\documentclass[apj]{emulateapj}
\usepackage{times}
\usepackage{natbib}
\usepackage[backref,breaklinks,colorlinks,citecolor=cyan]{hyperref} 
\usepackage{booktabs}
\usepackage{graphicx}
\usepackage{multirow}

\shorttitle{UDG Morphology}
\shortauthors{Mowla et al.}

\begin{document}

\title{Evidence of Absence of Tidal Features in the Outskirts of Ultra Diffuse Galaxies in the Coma Cluster}

\author{Lamiya Mowla\altaffilmark{1} Pieter van Dokkum\altaffilmark{1} Allison Merritt\altaffilmark{1} Roberto Abraham\altaffilmark{2} Masafumi Yagi\altaffilmark{3} Jin Koda\altaffilmark{4}}
\email{lamiya.mowla@yale.edu}

\altaffiltext{1}{Astronomy Department, Yale University, New Haven, CT 06511, USA}
\altaffiltext{2}{Department of Astronomy and Astrophysics, University of Toronto, 50 St. George Street, Toronto, ON M5S 3H4, Canada}
\altaffiltext{3}{Optical and Infrared Astronomy Division, National Astronomical Observatory of Japan, Mitaka, Tokyo, 181-8588, Japan}
\altaffiltext{4}{Department of Physics and Astronomy, Stony Brook University, Stony Brook, NY 11794-3800, USA}

\begin{abstract}
We study the presence of tidal features associated with ultra diffuse galaxies (UDGs) in galaxy clusters. Specifically, we stack deep Subaru images of UDGs in the Coma cluster to determine whether they show position angle twists at large radii. Selecting galaxies with central surface brightness $\mu(g,0) >24$ magarcsec$^{-2}$ and projected half-light radius $r_e>1.5$ kpc, we identify 287 UDGs in the \cite{Yagi2016} catalog of low surface brightness Coma objects. The UDGs have apparent spheroidal shapes with median S\'ersic index $\langle n\rangle =0.8$ and median axis-ratio $\langle b/a \rangle= 0.7$.  The images are processed by masking all background objects and rotating to align the major axis before stacking them in bins of properties such as axis ratio, angle of major axis with respect to the cluster center and separation from cluster center. Our image stacks reach further than 7 kpc ($\gtrsim4r_e$). Analysis of the isophotes of the stacks reveal that the ellipticity remains constant up to the last measured point, which means that the individual galaxies have a non-varying position angle and axis ratio and show no evidence for tidal disruption out to $\sim 4r_e$. We demonstrate this explicitly by comparing our stacks with stacks of model UDGs with and without tidal features in their outskirts. We infer that the average tidal radius of the Coma UDGs is $>$7 kpc and estimate that the average dark matter fraction within the tidal radius of the UDGs inhabiting the innermost 0.5 Mpc  of Coma is $>$ 99\%.
\end{abstract}

\keywords{galaxies: clusters: individual (Coma) --- galaxies: photometry --- galaxies: structure}

\section{Introduction}
\label{sec:introduction}

Ultra diffuse galaxies (UDGs) are extended low surface brightness galaxies with half-light radii comparable to those of $L_{\star}$ galaxies and stellar masses comparable to those of dwarf galaxies. UDGs were defined empirically as galaxies with central surface brightness $\mu(0,g)>24$ magarcsec$^{-2}$ and projected half-light radius (the radius containing 50\,\% of the light) $r_e>1.5$ kpc \citep{VanDokkum2015a}. Given their low stellar masses, it has been argued that UDGs may be analogs of other low mass galaxies, which have been puffed up by gravitational heating, tidal effects, or a high angular momentum \citep[e.g.,][]{Beasley2016a, Amorisco2016}. An alternative idea is that many are failed galaxies which quenched earlier than other galaxies of the same halo mass, through ram pressure stripping, stellar feedback driven galactic winds, or other effects \citep{VanDokkum2015,Yozin2015,Agertz2015,DiCintio2010,Beasley2016,Peng2016}.  Some UDGs have very high M/L ratios as determined from their kinematics \citep{VanDokkum2017,Beasley2016a}. However, whether the majority of UDGs form a stable population or whether many are part of a disrupting population of galaxies mixing in with the intracluster and intragroup light \citep{Searle1978} is still an outstanding question.

\begin{figure}[ht]
	\centering
	\includegraphics[width=0.48\textwidth]{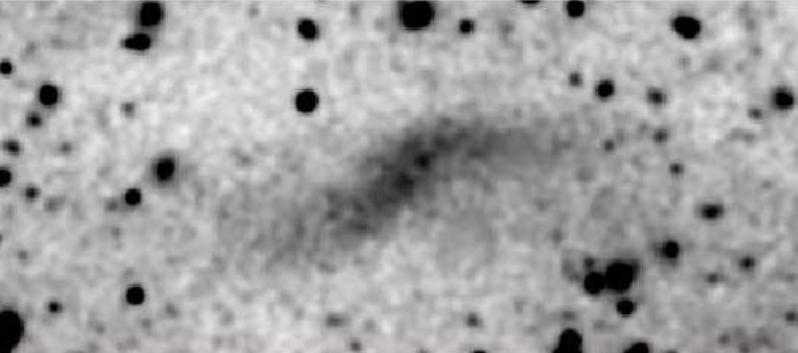}
	\caption{Example of a tidally disrupted galaxy (not a UDG) with an ``S"-shaped morphology. This is a co-added V and I image of the galaxy HCC-087 in the Hydra I cluster, taken from \cite{Koch2012}. The image covers 1.6'$\times$0.7' ($\sim$22$\times$10 kpc). }
	\label{fig:exampledistortion}
\end{figure}

As galaxies move through the cluster, they are disturbed due to the tidal field of the cluster and interactions with other galaxies. The effect is strongest beyond the tidal radius, where the gravitational force from the enclosed mass of the galaxy equals the gravitational force produced by the cluster. The tidal radius $r_t$ is given by

\begin{equation}
r_t = R \left( \frac{m}{3M}\right)^{1/3},
\label{eq:rt}
\end{equation}

where $R$ is the distance from the cluster center, $M$ is the total mass enclosed by the cluster within radius $R$, and $m$ is the total mass of the galaxy within radius $r_t$ (King 1962). Beyond the tidal radius stars can escape, giving rise to low surface brightness tidal tails in a characteristic ``S"-shaped pattern \citep[see][and Fig.\ \ref{fig:exampledistortion}]{Johnston2001, Bullock2005, Johnston2008,Koch2012}. Hence, the tidal radius can be determined by identifying the radius at which tidal tails appear, which in turn can be used to calculate the enclosed mass of the object. 

Imaging even the central regions of individual UDGs is challenging. This is demonstrated in Fig.\ 3 of van Dokkum et al.\ (2017), which shows deep HST images of UDGs in Coma: only the light within $\lesssim r_e$ is readily visible. An alternative to ultra-deep imaging is stacking a large number of galaxies with similar properties to greatly improve the overall depth reached at the cost of losing information of individual galaxies \citep{Zibetti2004,VanDokkum2010,Tal2011,DSouza2014}. The stacking technique requires deep images of a large number of galaxies with similar properties  \citep{Zibetti2004}. In this paper we are stacking deep $R$ band images of 231 UDGs in the Coma cluster from Suprime-Cam on the Subaru Telescope \citep{Koda2015,Yagi2016}. We use the fact that tidal tails produce a change in the position angle of isophotes to constrain their prominence. We simulate model galaxies with and without tidal tails to test their effect on stacked images, and compare them to UDG stacks. The radius at which tidal tail signatures appear allows us to estimate the observed average tidal radius of Coma UDGs and if no signatures are found we obtain a lower limit to the tidal radius. Using the tidal radius we can determine the average minimum enclosed mass of the stacked galaxies depending on their position in the cluster.

\section{Data and Sample Selection}
\label{sec:data}
	
\subsection{Data}

The cluster UDG sample for this study is a subsample of the 854 low surface brightness galaxies (LSB galaxies) found in the Coma cluster by \cite{Koda2015} using deep R band images, obtained with Suprime-cam on the Subaru telescope \citep{Okabe2014}. The selection criteria in that study were set such that all 47 of the Dragonfly UDGs were recovered, which included objects that are smaller than the canonical UDG criterion of $r_e>1.5$\,kpc.\footnote{In this paper we distinguish the generic term ``low surface brightness galaxies" and the term ``ultra diffuse galaxies" (UDGs). UDGs have $\mu(0,g)>24$ and $r_e>1.5$\,kpc. The Yagi et al.\ catalog contains many low surface brightness galaxies that are not UDGs according to this definition.} Details of the reduction procedure and how the objects were selected are given in \cite{Yagi2016}.  

\subsection{GALFIT fitting}	
\label{sec:mask}

We ran GALFIT \citep{Peng2009,Peng2002} on 201 pixel $\times$ 201 pixel (40."6 $\times$ 40."6) image cut-outs centered on the LSB galaxies to derive the physical parameters of the objects. All foreground and background objects, including point sources inside the LSB galaxies, were masked in the fits. Masks were created using the segmentation map from SExtractor \citep{Bertin1996} in two stages. In stage 1, we ran SExtractor on the individual stamps with a fixed detection threshold of 1.5 times the standard deviation above the background RMS level, 32 deblending sub-thresholds, and a minimum contrast parameter of 0.005. We also created FLAG maps for the images by flagging pixels at edges of frames. All objects in the segmentation map except the center of the LSB galaxy were masked in the primary mask. Using the primary mask as the bad pixel map, we used GALFIT to fit a one component S\'ersic profile to the LSB galaxy as an initial model. In stage 2 the residual map of the initial fit was used to create a second segmentation map with SExtractor with the deblending threshold raised to 64 and contrast parameter lowered to 0.001. This stage preferentially detects smaller and fainter objects that overlap with the (now-removed) LSB galaxy. The final mask is made by adding these two masks so that any bright objects other than the smooth component of the LSB galaxy are masked in the final image. Using this final mask as the bad pixel map, GALFIT is used to fit a single component S\'ersic profile with S\'ersic index and sky background allowed to vary. We judged the fits acceptable based on the goodness-of-fit $(\chi^2/\nu)<$1.2 and by visual inspection of the model fits.

Using the physical parameters from the GALFIT fit of each LSB galaxy, we select our UDG sample with the following definition of UDGs: $r_e>1.5$ kpc and $\mu(0,R)>$23.5 magarcsec$^{-2}$ \citep{VanDokkum2015}. The surface brightness cut was determined assuming a typical UDG color of $g-R\approx 0.5$ \citep[see][]{VanDokkum2015}. From the 857 LSB galaxy galaxies, 287 objects satisfy the UDG criterion and were selected to form our sample of Coma UDGs. 

\subsection{Properties of Coma UDGs}

\begin{figure*}[ht]
\centering
\includegraphics[width=0.48\textwidth]{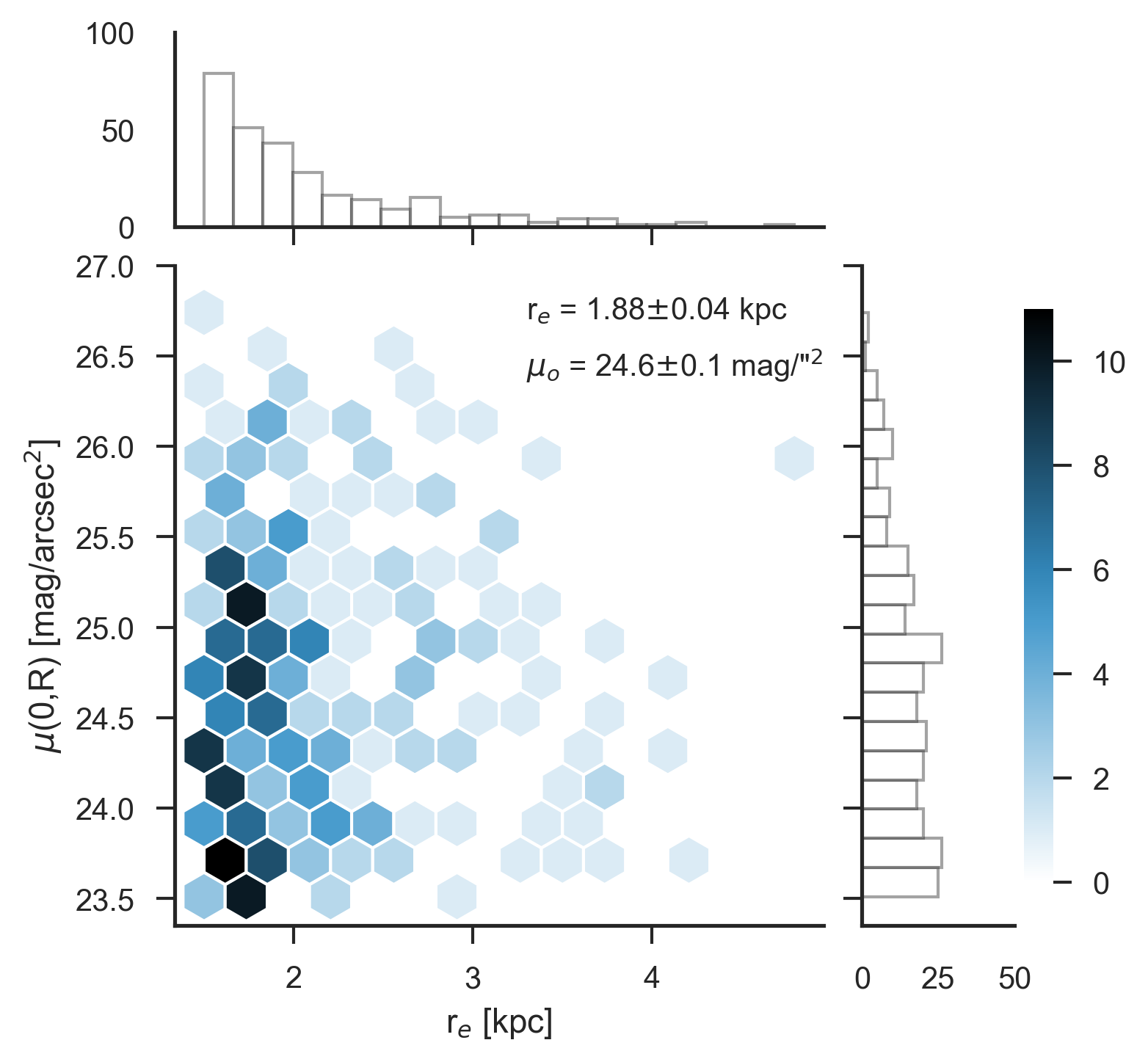}
\includegraphics[width=0.48\textwidth]{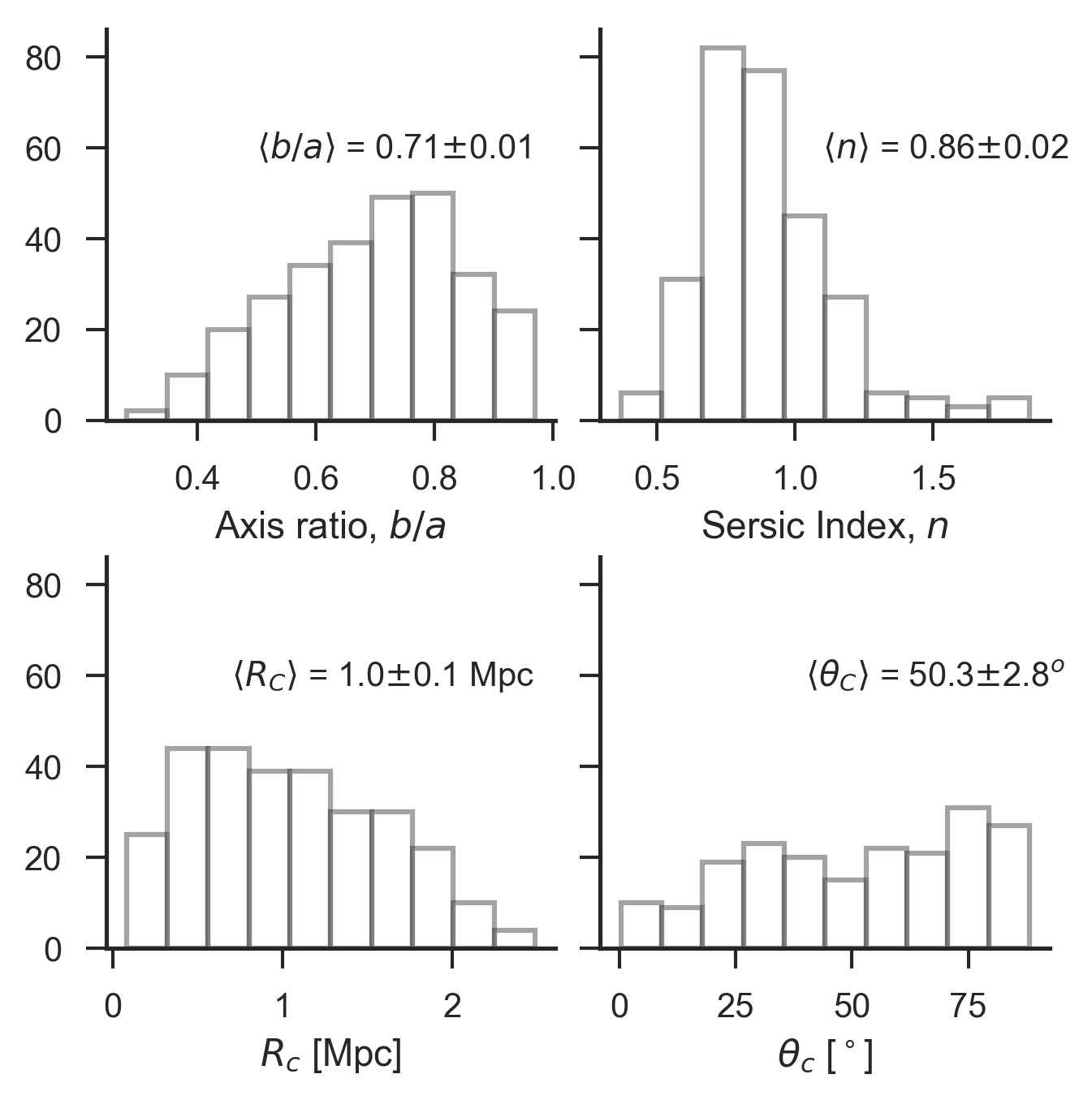}
\caption{Histograms of effective radii $r_e$, central surface brightness $\mu(R,0)$, axis ratio $b/a$, S\'ersic index $n$, angle between the major axis and the cluster-centric direction $\theta_C$, and distance from the cluster center $R_C$ ($\alpha = 12^h 59^m 35.^s7$ and $\delta = 27^o57"34'$). The \textit{left} plot shows the distribution of $r_e$ and $\mu(0,R)$ of the UDGs illustrating the selection criteria of $r_e>$1.5 kpc and $\mu(0,R)>$ 23.5 magarcsec$^{-2}$. The median values for each of the parameters and the bootstrap errors of the median values are shown on the plots. In the last histogram for the cluster-centric angle, UDGs whose $b/a<$0.8 are used only. Here $\theta_C$=0$^o$ means the major axis of UDG is aligned with the cluster-centric direction and $\theta_C$=90$^o$ means the minor axis is aligned with the cluster-centric direction.}
\label{fig:hist}
\end{figure*}

Figure \ref{fig:hist} shows the distribution of the physical parameters from our GALFIT fits of the UDGs in the Coma cluster. The median central surface brightness $\mu(0,R)$ of the UDGs is 24.6 $\pm 0.1$ magarcsec$^{-2}$, where the error is computed by bootstrap resampling. The median S\'ersic index is 0.88$\pm 0.02$ with 80$\%$ UDGs having $n$ between 0.6--1.1, in agreement with other studies \citep{Koda2015,VanderBurg2016,Roman2016}.
The median axis ratio is 0.71$\pm 0.01$, identical to the value found by \cite{Koda2015}, with a quarter of the UDGs having $b/a<$ 0.6. The median projected distance, $R_c$, of the UDGs from the cluster center is 1$\pm 0.1$ Mpc. There are fewer UDGs within 250 kpc of the cluster center; this may be an observational bias due to the brightness of the ICL or a true deficit, and is consistent with other UDG studies \citep{VanderBurg2016,Beasley2016a}. The UDGs near the center of the cluster are slightly rounder than those in the outskirts, with median $b/a =$ 0.75 for UDGs with $R_c <$ 0.65 Mpc and $b/a =$ 0.68 for UDGs with $R_c >$ 1.5 Mpc. The projected orientation of the major axis of UDGs relative to the cluster-centric direction is also shown in Figure \ref{fig:hist} where  $\theta_C$ is the angle between the major axis of the UDG and the line connecting the galaxy to the center of the cluster. Here, $\theta_c = 0^{o}$ means the major axis aligns in the cluster-centric direction while $\theta_c = 90^{o}$ means the minor axis aligns in cluster-centric direction. Again, a slightly bigger axis ratio is found for UDGs which are pointed towards the cluster center than those which tangential to it, with median $b/a =$ 0.78 for UDGs with $\theta_c <$ 25$^o$ and $b/a =$ 0.69 for UDGs with $\theta_c >$ 50$^o$. 

\section{Analysis}
\label{sec:stacking}

\subsection{Stacking}
The UDGs were divided in bins based on various properties, and the images of UDGs in each bin were processed and added together to create stacked images. For the stacking analysis, we used UDGs with $r_e <$ 2.5 kpc to ensure we are adding light from similar sized objects. The 231 UDGs which satisfy this criterion where visually inspected and used in the stacking analysis. The pre-stacking processing includes:

\begin{enumerate}
	\item shift the image to center the UDG on the central pixel
	\item rotate the image to align the major axis of the UDG with the y-axis
	\item apply the final mask created in the GALFIT stage (\ref{sec:mask}) to mask any foreground/background objects, including point sources on the UDG
	\item subtract the average sky count found by GALFIT from the image 
\end{enumerate}

The processed image has only the smooth component of the UDG light centered and vertically aligned along the major axis. The UDGs were binned in ranges of $b/a$, $\theta_C$ and $R_c$, to investigate any correlations of these properties with the presence of tidal features. We did not re-scale the sizes of the galaxies prior to stacking them but the images were normalized by the total flux from the GALFIT model. This normalization ensures that the stacks are not dominated by the brighter and larger UDGs. Stacks were created by adding together the processed images in each bin. The object masks were also summed to create weight maps, and the summed image was divided by its respective weight map to create averaged exposure-corrected stacks. The background subtraction is somewhat uncertain as the images show a very faint negative ``bowl" around all objects; this is a common artifact of over-subtraction of background during reduction of images of extended objects. We address this by applying a final background correction after stacking, by determining the median count at $R=8.5$\,kpc and subtracting this value. Effectively we set the galaxy flux to zero at this radius. This uncertainty in the background value is important for studies of the outer surface brightness profile, but is expected to have a negligible effect on the ellipticity profile.

\subsection{Mock Galaxy Images}

\begin{figure*}[ht]
\centering
\includegraphics[width=0.95\textwidth]{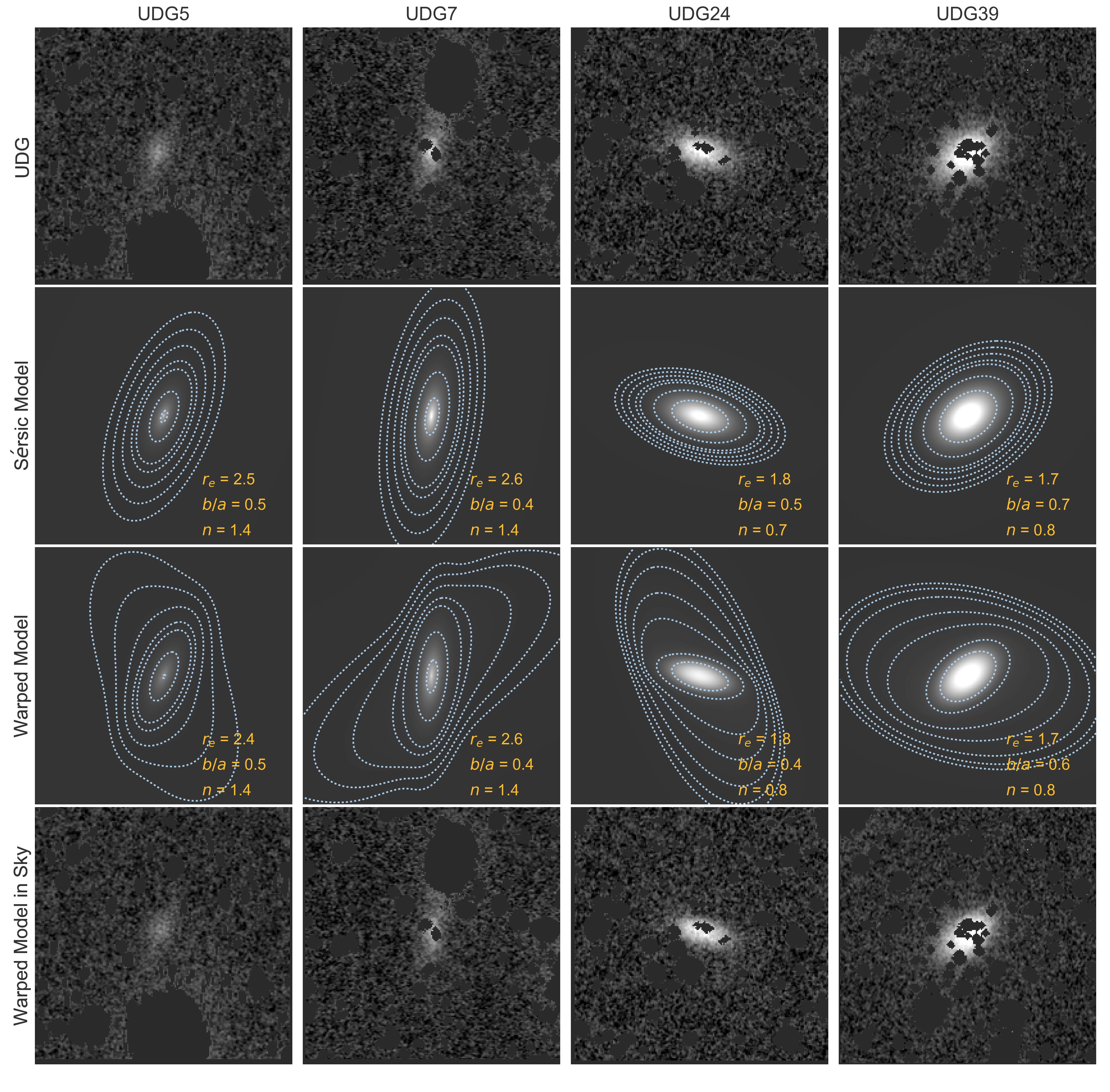}
\caption{Images of UDGs and their respective models with and without tidal tails. The top panels show Subaru $R$-band images of UDGs in the Coma cluster, with foreground and background objects including point sources inside the UDG masked. The names of the UDGs correspond to their IDs in the \cite{Yagi2016} catalog. The stamps span 40."6 $\times$ 40."6 (19kpc$\times$19kpc at the distance of Coma; scale length 0.47 kpcarcsec$^{-1}$). The second panel shows the S\'ersic models of UDGs as determined by GALFIT. The third panel shows the Warped model of galaxies with S-shaped light profile. The effective radius $r_e$, axis ratio $b/a$ and S\'ersic index $n$ from the GALFIT fits for each of the images are shown.The dotted contours show the isophotes. The bottom panel shows the Warped model of each UDG with their corresponding sky background and mask added for visual comparison with the UDG image in the top panel. } 
\label{fig:stamps}
\end{figure*}

We made mock images for the 231 UDGs to test the effect of tidal features in stacked images. Two types of model UDGs were created: i) S\'ersic, and ii) Warped.  The S\'ersic model is a simple ellipse with S\'ersic light profile and isophotes with constant axis ratio and position angle. This is the same as the GALFIT model of the galaxy. Simulations of tidally-distorted galaxies typically show breaks in the surface brightness profile where the tidal tails begin \citep[see][]{Johnston2001}. Our Warped model reproduces this feature along with the the axis ratios and position angles of the isophotes changing with radius. They are made by summing 100 ellipses with S\'ersic light profiles, with their axis ratio, S\'ersic index, central surface brightness  and position angle changing with radius to create S-shaped isophotes beyond $\sim$2 r$_e$. A hundred models are made for each UDG with a random combination of values for the starting $r_e$, $b/a$, $n$, $\mu_0$ within 0.5 times the true values of the respective parameters for the UDG. GALFIT is run on the hundred models and the model with fitted parameters that are closest to those of the observed UDG is chosen as the best Warped model. If a GALFIT parameter of the best-fit model is more than 20\,\% different from that of the observed UDG, then the model is rejected and the process is repeated by making 100 more models. The masks for each UDG are used as bad pixel maps during the GALFIT fit of Warped model so as to take into account the effect of masking on the fit. The end result is that we have a Warped model for each UDG that produces the same GALFIT parameters as were actually measured for that UDG, as illustrated in Fig.\ \ref{fig:stamps}. The direction of the warp is chosen at random, with the constraint that there are approximately equal numbers of UDG models with a warp in each direction. Details for creating images of Warped models are given in Appendix \ref{app:warp}. 

The bottom panel of Figure \ref{fig:stamps} shows the Warped model with their respective sky background and mask added. As can be seen GALFIT and our eyes cannot distinguish between the top and the bottom panel even though the underlying light profiles of the galaxies are significantly different. Due to the very low S/N of the UDGs methods such as fitting ellipses to trace the isophotes fail for individual UDGs. Conventional methods cannot trace the presence of tidal features of UDGs necessitating the use of stacking.

\subsection{Ellipse fitting}

The radial variation of the average structure of the galaxies is determined by fitting elliptical isophotes to the stacked images.
The fits are performed using the IRAF task $ellipse$, with logarithmic spacing of ellipses. The center position, axis ratio and position angle are allowed to vary with radius in the fits. 
	
\section{Results}
\label{sec:result}

\subsection{Stack of elongated UDGs}
\label{sec:stack_thin}
\begin{figure*}[ht]
	\centering
	\includegraphics[width=0.9\textwidth]{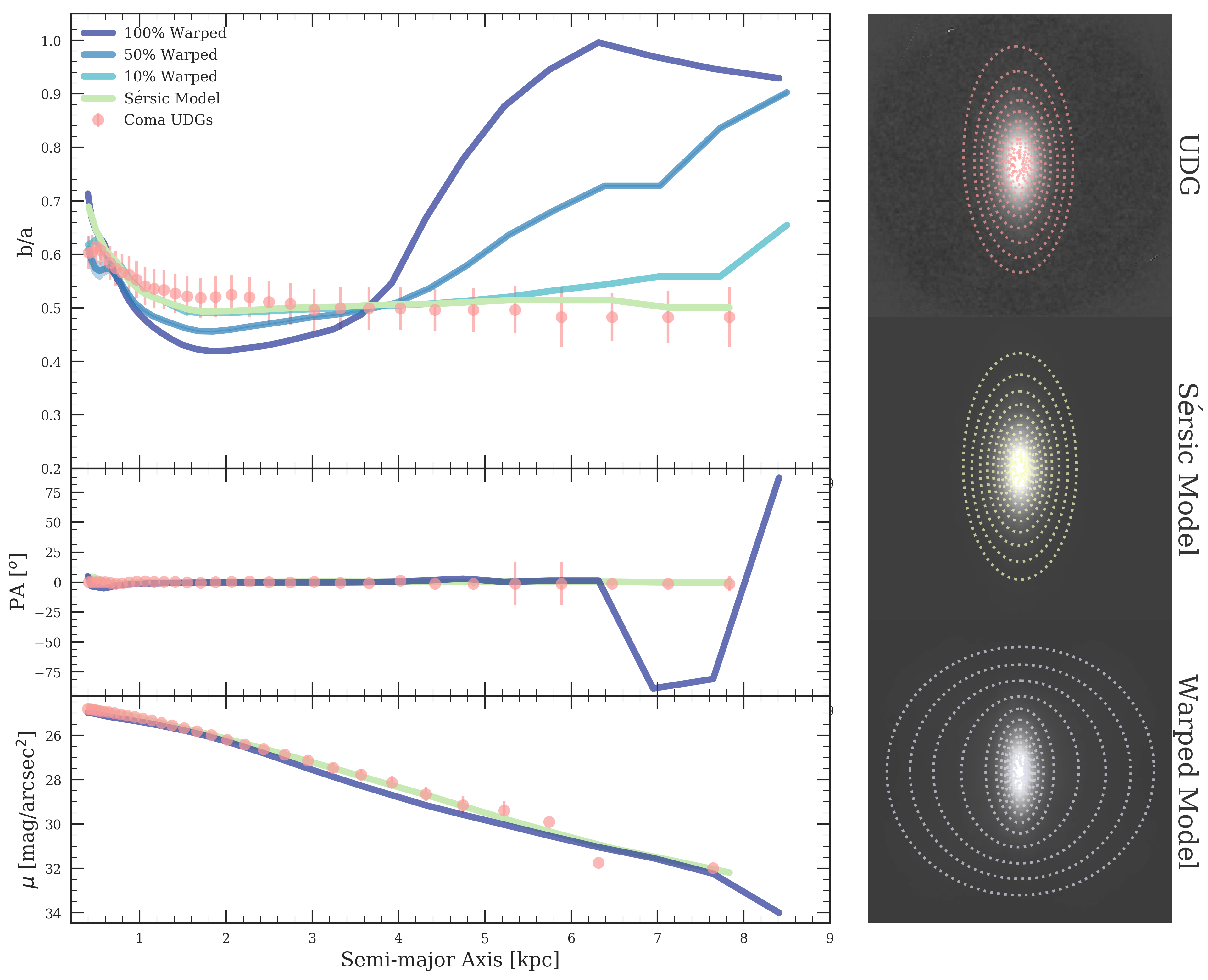}
	\caption{The axis ratio $b/a$, position angle $PA$ and surface brightness $\mu$ of the successive elliptical isophotes fitted on the stacked images with the IRAF task ELLIPSE are shown on the left panels. The stacked images are shown in the right panels, overplotted with their elliptical isophotes.}
	\label{fig:stac}
	\end{figure*}

The key result of our study is shown in the top panel of Fig.\ \ref{fig:stac}, which shows the observed axis ratio as a function of radius for the stack of the 68 UDGs with $b/a<0.6$ (pink points with errorbars). We focus on this subsample as the most elongated UDGs are the most sensitive to tidal features (as the fractional change in $b/a$ with radius is higher for UDG with smaller axis ratio). The axis ratio is nearly constant with radius out to the last measured point at $r\sim 8$\,kpc. This is remarkable, as it implies that the individual galaxies going into the stack are coherent, with the same axis ratio and position angle, out to this radius.

This is quantified and illustrated by comparing the observed stack to the two model stacks: the S\'ersic model, which is built from individual galaxies that have a constant ellipticity and position angle, and the Warped model, which is built from individual galaxies with ``S"-shaped tidal distortions.  The right panel shows the S\'ersic and Warped stacks overlaid with their isophotes from the ellipse-fitting analysis, while the left panel shows the axis-ratio, position angle and surface brightness of the elliptical isophotes.
As expected, the axis ratio of the stack of S\'ersic models remains nearly constant at $0.50\pm0.05$. However, the axis ratio of the stack of Warped models (shown by the 100\% Warped model line) increases from $0.50\pm0.05$ to $1.0\pm0.05$ between 4 kpc$<a<$8 kpc. This behavior is expected: when the S-shaped images of the Warped models  are added together in a stack, the resulting light profile has a circularized profile in the outskirts. Axis ratios of stacks with different fractions of S\'ersic and Warped models are also plotted to show the effect on the average axis ratio of the isophotes if only a small fraction of UDGs are disrupting. We compute the error bars in the axis ratio, position angle and surface brightness by bootstrap resampling of the stacks. The error bars for the models are shown as shaded region, which are barely visible due to their small sizes ($\sim 2\%$). For the 10$\%$ and 50$\%$ Warped stacks, we created 500 instances of stacks with different combinations of Warped and S\'ersic models and found $<3\%$ change in the results of the ELLIPSE analysis. This shows that the results are not dependent on the disruption of any particular galaxy. The stacked axis ratio of the observed UDGs closely follows the non-warped S\'ersic model out to the last measured point at $\sim 8$ kpc. In order to test whether larger UDGs were dominating the light in the outskirts from smaller UDGs, we also stacked the UDGs in bins of $r_e$ and found no significant change in axis ratio at larger radius in any of the size bins. Hence, our conclusion remains unchanged even when we add UDGs with $r_e >$2.5 kpc.

Note that all models reproduce the small scale variations in the ellipticity profile within $\sim 1.5$\,kpc.
This is because the PSF and the same masks have been applied to the models as to the observations. We tested this by creating stacks of models without any masks or PSF, with the resulting isophotes having perfectly smooth shapes.  The middle panel of Figure \ref{fig:stac} shows the position angles of the isophotes of the stacks. Since all of the UDGs and models are rotated such that their major axes are aligned with the y-axis, the average position angle of the stack maintains at zero for all the stacks. The change in the position angle of the Warped stacks beyond 7 kpc indicates that, at this radius, the strength of the warps causes the stack to be slightly more extended  along the (original) minor axis than along the major axis, as can be seen in the image of the Warped model.
The bottom panel shows the surface brightness of the isophotes.This has an almost exponential profile going from 24 magarcsec$^{-2}$ to $>30$ magarcsec$^{-2}$, although we caution that our straightforward background subtraction makes the interpretation difficult. 

\subsection{Stacking Subsamples of UDGs}

\begin{figure}[ht]
	\includegraphics[width=0.45\textwidth]{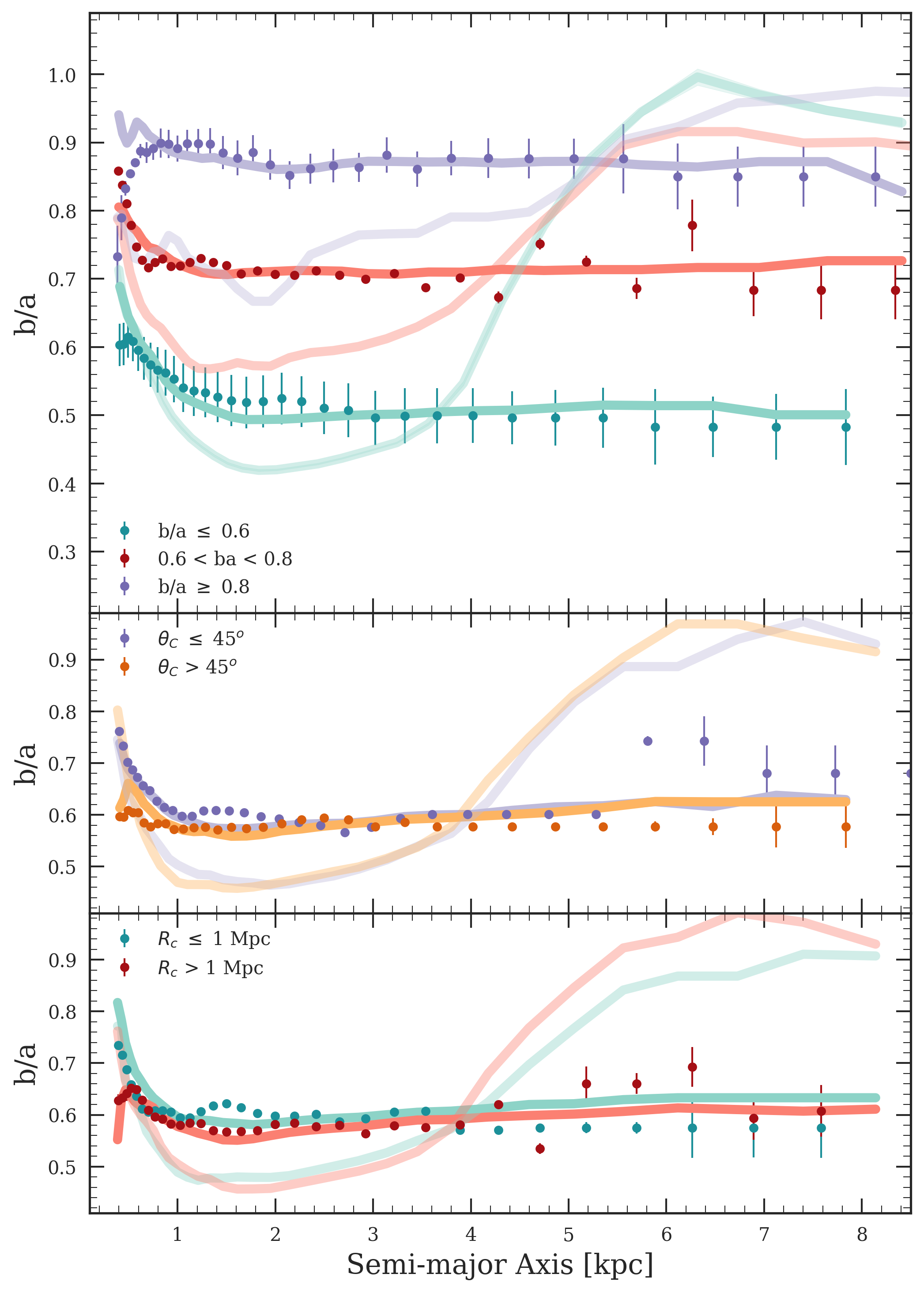}
	\caption{The axis ratios of the isophotes of the different stacks made in bins of different properties:  axis ratio (top panel), angle of the major axis with respect to the direction toward the cluster center (middle panel) and distance from the cluster center (bottom panel). The dots show the UDG data, the darker line shows the S\'ersic model and the lighter line shows the Warped model. With the possible exception of galaxies with $\theta_c<45^o$, the axis ratio remains constant with radius in all subsamples.}
	\label{fig:ba}
\end{figure}

\begin{deluxetable}{c | c c c}[ht]
\tablewidth{0.45\textwidth}
\tablecaption{Number of UDGs in each stack}
\tablehead{\colhead{Property} & \colhead{Bin} & \colhead{Number of UDGs} & \colhead{$\langle b/a \rangle$}\tablenotemark{a}}
\startdata
\\
\multirow{3}{*}{\parbox{2cm}{\centering Axis ratio}} & $b/a \leq 0.6$ & 68 & 0.50 $\pm$ 0.07 \\
& $0.6<b/a<0.8$ & 85 & 0.71 $\pm$ 0.06\\
& $b/a \geq 0.8$ & 78 & 0.86 $\pm$ 0.07 \\
\\
\hline
\\
\multirow{2}{*}{\parbox{2cm}{\centering Cluster-centric angle\tablenotemark{b}}} & $\theta_c\leq 45^o$ & 62 & 0.59 $\pm$ 0.15\\
& $\theta_c>45^o$ & 70 & 0.59 $\pm$ 0.14 \\
\\
\hline
\\
\multirow{2}{*}{\parbox{2cm}{\centering Distance from cluster-center\tablenotemark{b}}} & $R_c \leq 1.0$ Mpc & 67 & 0.63 $\pm$ 0.12 \\
& $R_c>1.0$ Mpc & 67 & 0.59 $\pm$ 0.13  \\
\enddata
\tablenotetext{a}{Median axis-ratio of the stack}
\tablenotetext{b}{Only UDGs with $b/a<$0.75 used in the stack}


\tablerefs{}
\label{tab:sub}
\end{deluxetable}

In section \ref{sec:stack_thin} we stacked the UDGs with the smallest axis ratios, finding no evidence for tidal distortions. Here we analyze other subsamples of UDGs, to determine whether there is an increase in the stacked axis ratio with radius for any of them.  Table \ref{tab:sub} shows the number of UDGs in each sub-sample stack and the median axis-ratio of each stack. We first consider other axis ratio bins, that is, the average radial trend of the axis ratio at large radii for galaxies with different GALFIT-determined axis ratios in their inner parts. The top panel of Figure \ref{fig:ba} shows the radial behavior of the axis ratio for four stacks with different GALFIT-determined axis ratio. The dots with error bars show the UDG stack, the darker line shows the S\'ersic model stack and the light line shows the Warped model stack. Each stack contains 60-85 UDGs. As expected, for the S\'ersic models, the axis ratio of the isophotes remain constant at the median axis ratio of the stack while the Warped model axis ratio rises at $>$4kpc in all three stacks, with the effect being the strongest for the thinnest UDGs. The observed axis ratio closely follows the S\'ersic model stack except for the $0.6<b/a<0.8$ bin, which has some irregularity beyond radius $>4.5$ kpc but follows the S\'ersic model within two standard deviations. The median residual is for the S\'ersic model (-0.01) is smaller than the median residual for the Warped model (-0.10). 

\begin{figure*}[ht]
	\centering
	\includegraphics[width=0.8\textwidth]{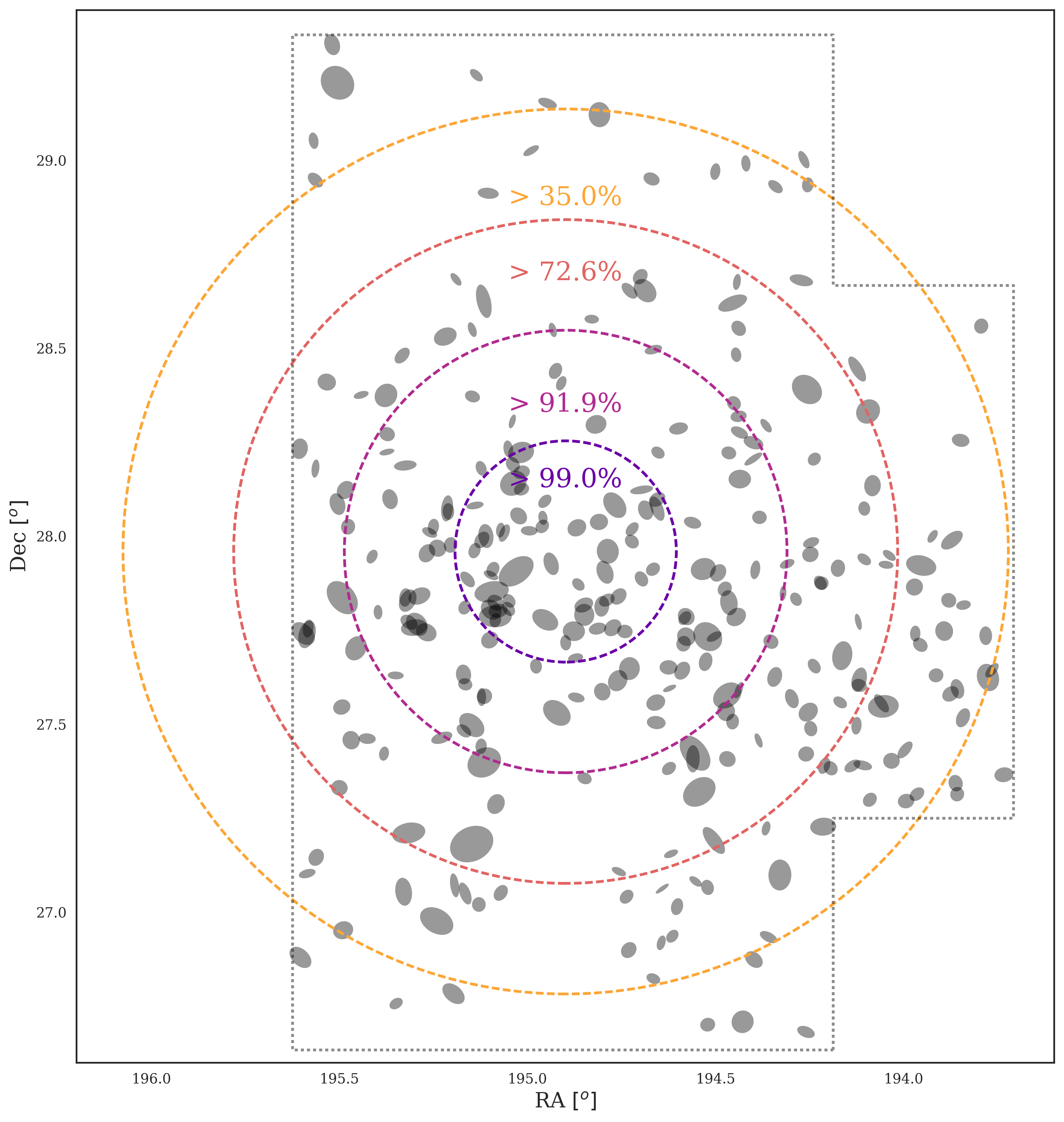}
	\caption{Spatial distribution of UDGs in the Coma cluster. The axis ratio and relative sizes of the ellipses represent the GALFIT fit values of our UDG sample (with the symbol size magnified by a factor of 20). The dotted grey line trace the field covered by the Subaru imaging with a total area of 4.1 deg$^2$ \citep{Okabe2014}. The dashed circles represent the 0.5, 1.0, 1.5 and 2.0 Mpc separation from the cluster center at the distance of Coma cluster. Using our estimated tidal radius of 7 kpc, the total mass enclosed within the tidal radius is calculated using equation \ref{eq:m} for each of the separation bins marked by the dashed lines. From the calculated enclosed mass and assuming stellar mass $M_{star} \approx 1.1 \times 10^8 M_{\odot}$, the minimum dark matter required in order for the UDG not be tidally disrupted is estimated. The minimum dark matter fractions for the UDGs within each separation bin are shown on the plot.}
	\label{fig:map}
\end{figure*}

Next we test whether UDGs whose major axes are (mis-)aligned with the cluster-centric direction are preferentially tidally disrupted. This is parameterized by the projected angle, $\theta_c$, between the major axis of the UDG and the line connecting the UDG with the center of the cluster. For UDGs whose major axis points towards the center of the cluster $\theta_C=0$ (for $\theta_c=90^o$ the major axis is perpendicular to the direction toward the cluster center, and the minor axis points to the cluster center). For each of the stacks, only galaxies with $b/a <0.7$ are included so that the position angle of the major axis is well-determined. The UDGs with major axis aligned with the cluster center ($\theta_C<45^o$) are in purple and those with minor axis aligned with cluster center ($\theta_C>45^o$) are in orange. Both the stacks have similar median axis-ratio with the UDGs pointed towards the cluster center with $b/a=0.6\pm0.03$ and the UDGs tangential with the cluster center with $b/a=0.59\pm0.03$. The slight irregularity beyond $>4.5$ kpc for the $\theta_C>45^o$ stack as well as for the $0.6<b/a<0.8$ stack from the above panel can be attributed to distortions in the galaxies which are not modelled by our S-shaped Warped model. 

Finally, we determine whether UDGs that are closer to the center of the cluster are more disrupted than those at larger distances. This might be expected as tidal forces are stronger near the center than in the outskirts.  The bottom panel of Figure \ref{fig:ba} shows the stacks of the UDGs with the smaller projected distance to the cluster center ($R_c<1.0$ Mpc in green) and those with largest projected distance from the cluster center ($R_c>1.0$ Mpc in red). Again a slight irregularity beyond $>4.5$ kpc is noticed for the $R_c>1$ Mpc stack showing presence of some distortion, but overall both the stacks follow the S\'ersic model stack. We note that $\theta_c$ and $R_c$ are projected properties; trends with the physical distance of a galaxy to the center of the cluster, or the 3D vector connecting a galaxy to the cluster center, will necessarily be somewhat diluted in projection.

\section{Discussion and Conclusions}

We have demonstrated that the stacking technique can recover simulated tidal features: models with an artificially-induced S-shaped tidal distortion beyond $\sim 2r_e$ (4\,kpc) show a pronounced increase in the observed axis ratio of the stack (see Fig.\ 4). We selected UDGs with $r_e<$2.5 kpc for our stacking analysis. The increase in axis ratio in stacked images is not seen in the data: the axis ratio profile remains flat out to the last measured point at $\sim 7.5$\,kpc, or $>$ 4 $<$$r_e$$>$. We conclude that the tidal radius of the majority of Coma UDGs likely lies beyond this point, and determine a lower limit of $r_t\gtrsim 7$ kpc for most UDGs.


With our estimated tidal radius of the majority of galaxies we can now calculate the average enclosed mass within the tidal radius, depending on their position in the cluster. Rewriting Eq.\ref{eq:rt},

\begin{equation}
m = 3M \left( \frac{r_t}{R} \right)^3
\label{eq:m}
\end{equation}

the enclosed mass of the UDGs within the tidal radii can be estimated, where $M=1.88\times10^{15}h^{-1}M_{\odot}$ and $R=1.99 h^{-1}$ Mpc, the virial mass and the virial radius of the Coma cluster respectively \citep{Kubo2007}. This means for an UDG at distance of 0.5 Mpc from the cluster center the mass enclosed within the tidal radius $\gtrsim 1.6\times10^{10}$M$_\odot$. The median absolute \textit{R} band magnitude $\langle M_R \rangle = 14.1$ and we estimate the median stellar mass of all UDGs $\langle M_{star}\rangle \sim 1.1 \times 10^8 M_{\bigodot}$, calculated by scaling to the stellar mass estimate of the Coma cluster UDGs in \cite{VanDokkum2015}. The dark matter fraction of these innermost UDGs within is then $\gtrsim$99\% within 7\,kpc. Since our stacks of galaxies in bins of distance from the cluster center did not show any sign of tidal tail for the UDGs closest to the cluster center, we are assuming that these central UDGs also have tidal radius $\gtrsim$7 kpc. Figure \ref{fig:map} shows the minimum dark matter fraction of UDGs in the Coma cluster depending on their position in the cluster. 
The high dark matter fractions that we derive are consistent with estimates based on dynamics \citep{Beasley2016,VanDokkum2017} and globular cluster counts \citep{Peng2016,Beasley2016,VanDokkum2017}.

Irrespective of the quantitative constraints on the dark matter fraction, the qualitative conclusion from this study is that we can rule out that most cluster UDGs resemble objects such as HCC-087 (Fig.\ref{fig:exampledistortion}). This result does not rule out that a small fraction of the cluster UDGs might be undergoing disruption, or may have features which are not modelled by our S-shaped warp, similar to those UDGs observed in the field \citep{Merritt2016,Roman2016}.

We note here that our analysis is only sensitive to S-shaped distortions, and we cannot rule out other types of tidal features.
Deeper data may provide better constraints for individual galaxies, although sky subtraction issues may limit the analysis unless great care is taken to obtain an optimally-flat background in the observations.

\acknowledgments
We thank the anonymous referee for a very constructive and helpful report. Support from NSF grant AST-1312376 is gratefully acknowledged.

\bibliographystyle{apj.bst}
\bibliography{udg_stack.bib}

\begin{thebibliography}{}
\expandafter\ifx\csname natexlab\endcsname\relax\def\natexlab#1{#1}\fi

\bibitem[{{Agertz} \& {Kravtsov}(2016)}]{Agertz2015}
{Agertz}, O., \& {Kravtsov}, A.~V. 2016, \apj, 824, 79

\bibitem[{Amorisco \& Loeb(2016)}]{Amorisco2016}
Amorisco, N.~C., \& Loeb, A. 2016, 5

\bibitem[{{Beasley} {et~al.}(2016){Beasley}, {Romanowsky}, {Pota}, {Navarro},
  {Martinez Delgado}, {Neyer}, \& {Deich}}]{Beasley2016}
{Beasley}, M.~A., {Romanowsky}, A.~J., {Pota}, V., {et~al.} 2016, \apjl, 819,
  L20

\bibitem[{Beasley \& Trujillo(2016)}]{Beasley2016a}
Beasley, M.~A., \& Trujillo, I. 2016, The Astrophysical Journal, Volume 830,
  Issue 1, article id. 23, 6 pp. (2016)., 830, 2

\bibitem[{Bertin \& Arnouts(1996)}]{Bertin1996}
Bertin, E., \& Arnouts, S. 1996, Astronomy and Astrophysics Supplement Series,
  117, 393

\bibitem[{Bullock \& Johnston(2005)}]{Bullock2005}
Bullock, J.~S., \& Johnston, K.~V. 2005, The Astrophysical Journal, 635, 931

\bibitem[{{Di Cintio} {et~al.}(2016){Di Cintio}, Brook, Dutton, {Macc{\`{i}}
  O}, Obreja, \& Dekel}]{DiCintio2010}
{Di Cintio}, A., Brook, C.~B., Dutton, A.~A., {et~al.} 2016, MNRAS, 000, 1

\bibitem[{D'Souza {et~al.}(2014)D'Souza, Kauffman, Wang, \&
  Vegetti}]{DSouza2014}
D'Souza, R., Kauffman, G., Wang, J., \& Vegetti, S. 2014, Monthly Notices of
  the Royal Astronomical Society, 443, 1433

\bibitem[{Johnston {et~al.}(2008)Johnston, Bullock, Sharma, Font, Robertson, \&
  Leitner}]{Johnston2008}
Johnston, K.~V., Bullock, J.~S., Sharma, S., {et~al.} 2008, The Astrophysical
  Journal, 689, 936

\bibitem[{{Johnston} {et~al.}(2002){Johnston}, {Choi}, \&
  {Guhathakurta}}]{Johnston2001}
{Johnston}, K.~V., {Choi}, P.~I., \& {Guhathakurta}, P. 2002, \aj, 124, 127

\bibitem[{Koch {et~al.}(2012)Koch, Burkert, Rich, Collins, Black, Hilker, \&
  Benson}]{Koch2012}
Koch, A., Burkert, A., Rich, R.~M., {et~al.} 2012, arXiv:arXiv:1207.2762v1

\bibitem[{Koda {et~al.}(2015)Koda, Yagi, Yamanoi, \& Komiyama}]{Koda2015}
Koda, J., Yagi, M., Yamanoi, H., \& Komiyama, Y. 2015, arXiv:1506.01712

\bibitem[{Kubo {et~al.}(2007)Kubo, Stebbins, Annis, Dell'Antonio, Lin,
  Khiabanian, \& Frieman}]{Kubo2007}
Kubo, J.~M., Stebbins, A., Annis, J., {et~al.} 2007, arXiv:0709.0506

\bibitem[{Merritt {et~al.}(2016)Merritt, van Dokkum, Danieli, Abraham, Zhang,
  Karachentsev, \& Makarova}]{Merritt2016}
Merritt, A., van Dokkum, P., Danieli, S., {et~al.} 2016, The Astrophysical
  Journal, Volume 833, Issue 2, article id. 168, 12 pp. (2016)., 833,
  arXiv:1610.01609

\bibitem[{Okabe {et~al.}(2014)Okabe, Futamase, Kajisawa, \&
  Kuroshima}]{Okabe2014}
Okabe, N., Futamase, T., Kajisawa, M., \& Kuroshima, R. 2014, The Astrophysical
  Journal, 784, 90

\bibitem[{Peng {et~al.}(2002)Peng, Ho, Impey, \& Rix}]{Peng2002}
Peng, C.~Y., Ho, L.~C., Impey, C.~D., \& Rix, H.-W. 2002, arXiv:0204182

\bibitem[{Peng {et~al.}(2009)Peng, Ho, Impey, \& Rix}]{Peng2009}
---. 2009, arXiv:0912.0731

\bibitem[{{Peng} \& {Lim}(2016)}]{Peng2016}
{Peng}, E.~W., \& {Lim}, S. 2016, \apjl, 822, L31

\bibitem[{{Rom{\'a}n} \& {Trujillo}(2017)}]{Roman2016}
{Rom{\'a}n}, J., \& {Trujillo}, I. 2017, \mnras, 468, 703

\bibitem[{Searle \& Zinn(1978)}]{Searle1978}
Searle, L., \& Zinn, R. 1978, Astrophysical Journal

\bibitem[{Tal \& van Dokkum(2011)}]{Tal2011}
Tal, T., \& van Dokkum, P.~G. 2011, The Astrophysical Journal, 731, 89

\bibitem[{van~der Burg {et~al.}(2016)van~der Burg, Muzzin, \&
  Hoekstra}]{VanderBurg2016}
van~der Burg, R. F.~J., Muzzin, A., \& Hoekstra, H. 2016, arXiv:1602.00002

\bibitem[{van Dokkum {et~al.}(2010)van Dokkum, Whitaker, Brammer, Franx, Kriek,
  Labb{\'{e}}, Marchesini, Quadri, Bezanson, Illingworth, Muzzin, Rudnick, Tal,
  \& Wake}]{VanDokkum2010}
van Dokkum, P., Whitaker, K., Brammer, G., {et~al.} 2010, $\backslash$Apj, 709,
  1018

\bibitem[{{van Dokkum} {et~al.}(2016){van Dokkum}, {Abraham}, {Brodie},
  {Conroy}, {Danieli}, {Merritt}, {Mowla}, {Romanowsky}, \&
  {Zhang}}]{VanDokkum2017}
{van Dokkum}, P., {Abraham}, R., {Brodie}, J., {et~al.} 2016, \apjl, 828, L6

\bibitem[{van Dokkum {et~al.}(2015{\natexlab{a}})van Dokkum, Abraham, Merritt,
  Zhang, Conroy, \& Geha}]{VanDokkum2015a}
van Dokkum, P.~G., Abraham, R., Merritt, A., {et~al.} 2015{\natexlab{a}},
  Astrophysical Journal, 804, 1

\bibitem[{van Dokkum {et~al.}(2015{\natexlab{b}})van Dokkum, Romanowsky,
  Abraham, Brodie, Conroy, Geha, Merritt, Villaume, \& Zhang}]{VanDokkum2015}
van Dokkum, P.~G., Romanowsky, A.~J., Abraham, R., {et~al.} 2015{\natexlab{b}},
  Astrophysical Journal, 804, 1

\bibitem[{Yagi {et~al.}(2016)Yagi, Koda, Komiyama, \& Yamanoi}]{Yagi2016}
Yagi, M., Koda, J., Komiyama, Y., \& Yamanoi, H. 2016, The Astrophysical
  Journal Supplement Series, 225, 11

\bibitem[{Yozin \& Bekki(2015)}]{Yozin2015}
Yozin, C., \& Bekki, K. 2015, 8, 1

\bibitem[{Zibetti {et~al.}(2004)Zibetti, White, \& Brinkmann}]{Zibetti2004}
Zibetti, S., White, S. D.~M., \& Brinkmann, J. 2004, Mon. Not. R. Astron. Soc,
  347, 556

\end{thebibliography}

\appendix

\section{Constructing the Warped models}
\label{app:warp}

\begin{figure}[h]
	\centering
	\includegraphics[width=0.7\textwidth]{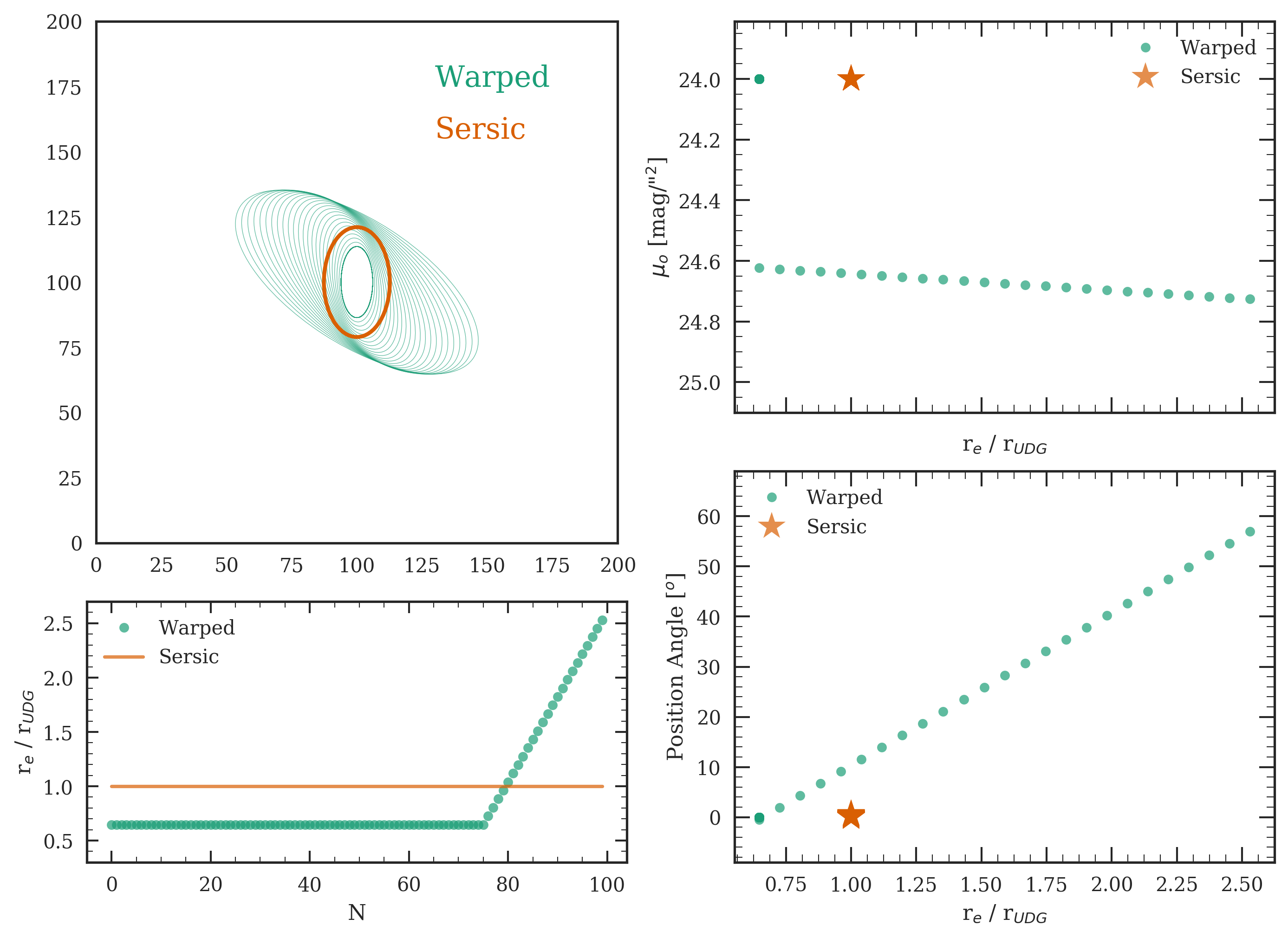}
	\caption{Making S\'ersic and Warped model with overlapping ellipses. The left panel shows the position of the ellipses making up the S\'ersic and the Warped model (top) and the change in the effective radius of the consecutive ellipses ($r_e$) normalized by the effective radius of the UDG modeled ($r_{UDG}$) (bottom). The models are rotated by PA$_{UDG}$ to align the major axis with the y-axis. Each model is made up of a 100 ellipses, with S\'ersic light profiles having the same center. The effective radii of the ellipses in the S\'ersic model remain constant at the effective radius of the UDG ($r_{UDG}$) being modeled The Warped models are created by adding an inner component with a fixed position angle to an outer component with  position angles varying with radius. The inner component has a fixed effective radius ($r_{in}$) smaller than $r_{UDG}$, while the outer component has effective radius ($r_{out}$) that increases with N from $r_{in}$ to $\sim 5r_{in}$. The right panel shows central surface brightness (top) and position angle (bottom) of the ellipses as a function of normalized effective radius of the ellipses $r_e/r_{UDG}$. In the S\'ersic model all the ellipses have the same central surface brightness and position angle, equal to the value of the UDG being modeled For the Warped model, the inner component have the same central surface brightness and position angle as the UDG. For the outer component, the ellipses get fainter and their position angle increases by up to 60$^o$.}
	\label{fig:modmake}
\end{figure}

\begin{figure}[h]
	\centering
	\includegraphics[width=0.8\textwidth]{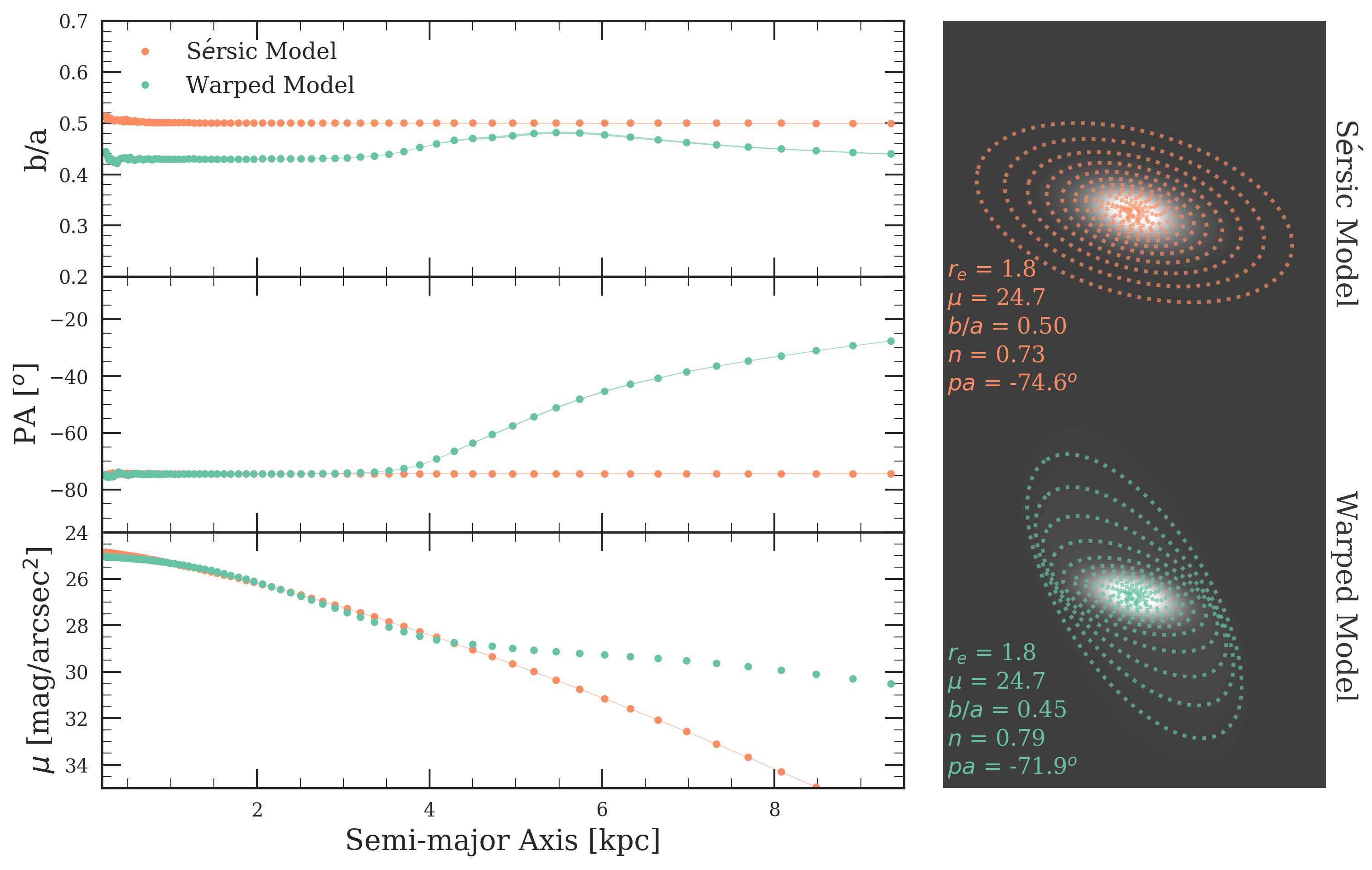}
	\caption{ELLIPSE analysis on individual S\'ersic and Warped models. The axis ratio ($b/a$) and position angle (PA) is constant and the surface brightness falls smoothly for the S\'ersic model. However, a distortion at $r>2r_e$ is noticed in all three parameters for the Warped model. This feature is observed in brighter warped galaxies as well as in disrupted satellite galaxies in numerical simulations \citep[see][]{Johnston2001}. Due to the low S/N of UDGs, such analysis cannot be done directly on observations.}
	\label{fig:modellipse}
\end{figure}

\begin{figure*}[h]
	\centering
	\includegraphics[width=\textwidth]{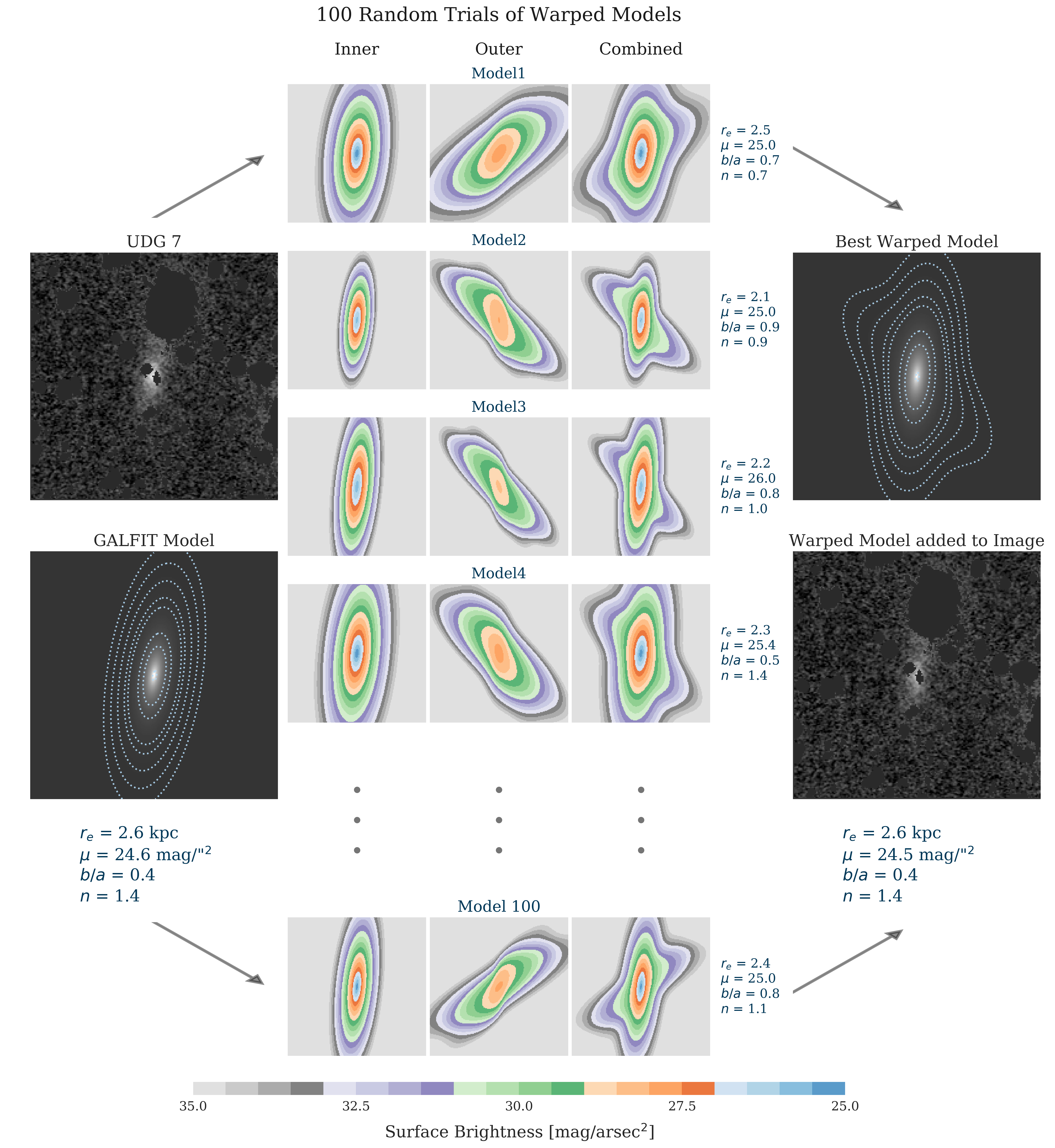}
	\caption{The making of the Warped models. The left panel shows a masked UDG image (top) and the S\'ersic model fit by GALFIT (bottom). A hundred Warped models are made for each UDG, each model consisting of an inner component and a warped outer component as shown in the middle panel. GALFIT is run on all 100 models and the model with the least difference in values of $r_e$, $\mu_0$, $b/a$ and $n$ between the UDG and the model is chosen. The right panel shows the best fit Warped model (top) and the Warped model added to the masked image of the sky (bottom).}
	\label{fig:app_warp}
\end{figure*}

The Warped models are created by adding an inner ellipsoidal model with a fixed position angle to an outer ellipsoidal model whose position angle varies with radius.  A Warped model is created for each UDG by adding the inner and outer components with the same center, both of which follow S\'ersic light profiles. For each model, we require that the structural parameters, as derived from a single-component fit by GALFIT, are identical to those that are actually measured from the image. This is achieved by creating 100 randomly generated models, and choosing the one whose fitted parameters are closest to those derived from the actual UDG image.  The parameters we match between the UDG and the model are: effective radius ($r_{UDG}$), central surface brightness in $R$ band ($\mu_{UDG}$), S\'ersic index ($n_{UDG}$) and axis-ratio ($b/a_{UDG}$). The direction of the warp of the best-fit model is chosen at random, with the constraint that there are approximately equal numbers of models with warp in each direction.

The inner component is made by superimposing ellipses at the center. All the ellipses have the same $r_e$, $\mu_0$, $b/a$ and $n$ and their values are chosen randomly with the following constraints (parameter subscript ``UDG" refers to the parameter value for the UDG being modelled):
\begin{enumerate}
\item Inner effective radius: $0.6 r_{UDG}<r_{in}<r_{UDG}$
\item Inner central surface brightness: $\mu_{in} = \mu_{UDG}$
\item Inner S\'ersic index: $0.6 n_{UDG}<n_{in}<n_{UDG}$
\item Inner axis ratio: $0.6 b/a_{UDG}<b/a_{in}<b/a_{UDG}$
\item Position angle of the inner ellipses with respect to the major axis of the UDG is zero.
\end{enumerate}

The outer component is made by superimposing ellipses at the center as well, but the various parameters of the ellipses change gradually:
\begin{enumerate}
    \item Outer effective radius changes linearly from $r_{in}$ to $\sim 3r_{in}$, with the radius changing by $\sim$0.15 kpc/ellipse.
    \item Outer central surface brightness: the flux of the center of the ellipse drops off as
    \begin{equation}
    \mu_{out}=\mu_{in}-2.5 \log_{10}(a\exp(-bN))
    \end{equation}
    where, $a$ and $b$ are random numbers in the range $0.1<a<0.9$ and $0.01<b<0.05$ respectively and are allowed to vary between different instances of the model. N refers to the index of the ellipse where N$=$1 is the innermost ellipse and N$=$100 is the outermost ellipse.
    \item Outer S\'ersic index: $0.2 n_{in}<n_{out}<n_{in}$
    \item Outer axis ratio: $b/a_{out} = b/a_{in}$
    \item Position angle of the ellipses with respect to the major axis increases linearly from 0$^o$ to $\sim$60$^o$ at a rate of $\sim$3$^o$/ellipse. 
\end{enumerate}

The inner and outer components are averaged together to produce the combined image with S-shaped outer isophotes (as shown in Fig. \ref{fig:modmake}), akin to the image of Fig.\ 1. After generating the models they are fit with a single component S\'ersic profile using GALFIT, and the fit values of $r_e$, $\mu_0$, $b/a$ and $n$ are compared with the values that are measured from the image of the UDG. The model with the smallest difference is chosen as the best Warped model for that UDG. If the best model has one parameter deviating by $>20\%$ from the original value, the whole set of 100 is thrown out and the process is repeated by creating 100 more models until a best-fit model is found where all the parameters deviate by $<20\%$.

We do not add any noise to these models but do apply their respective masks on them to take into account the effect of masking on the stacked images. 

\end{document}